\documentclass[a4paper,10pt]{article}
\usepackage{pos}
\usepackage{subcaption}
\usepackage{color,soul}
\usepackage{float}

\usepackage{braket}
\usepackage{blkarray}
\usepackage{url}

\title{Real time evolution and a traveling excitation in SU(2) pure gauge theory on a quantum computer}

\author{Sarmed A~Rahman}
\author{Randy Lewis}
\author*{E.~Mendicelli}
\author{Sarah Powell}

\affiliation{Department of Physics and Astronomy, York University,\\ Toronto, ON, M3J 1P3, Canada}

\emailAdd{emanuelemendicelli@hotmail.it}

\abstract{The Hamiltonian approach can be used successfully to study the real-time evolution of a non-Abelian lattice gauge theory on the available noisy quantum computers. In this work, results from the real-time evolution of SU(2) pure gauge theory on IBM hardware are presented. The long real-time evolution spanning dozens of Trotter steps with hundreds of CNOT gates and the observation of a traveling excitation on the lattice were made possible by using a collection of error mitigation techniques. Self-mitigation is our novel tool, which consists of using the same physics circuit as a noise-mitigation circuit.}

\FullConference{%
The 39th International Symposium on Lattice Field Theory,\\
8th-13th August, 2022,\\
Rheinische Friedrich-Wilhelms-Universität Bonn, Bonn, Germany
}

\begin{document}
\maketitle

\section{Introduction and motivations}
The quantum computers available in the current noisy intermediate-scale quantum (NISQ) era \citep{JohnPreskillNISQ} represent a prototype of the future in which the quantum computer will be as fault-tolerant as the classical computer \citep{Aharonov_et_all}. Conceptually the quantum computer offers interesting possibilities for the computation of lattice gauge theories as explained by the authors of \citep{Bauer:2022hpo}. Compared to a classical computer where the storage of the state vectors scales exponentially, the scaling on a quantum computer can be polynomial. Polynomial scaling makes it computationally feasible to use the Hamiltonian formulation in place of the common Lagrangian formulation, which requires the use of a Monte Carlo method. Furthermore,  the Hamiltonian formalism gives direct access to the real-time dynamics of the theory, and it is completely free of the infamous sign problem \citep{Zi-Xiang_Hong,Sign_problem_Gattringer} that affects the Monte Carlo method, opening the possibility of a rigorous study of theories with non-zero baryon density.\\

However the use of the current quantum hardware is not free from challenges. In fact, the quantum resources are limited in terms of number of qubits and their connectivity with each other. The available gates, in particular the CNOT gate, are noisy due to their imperfect realization leading to errors in the measurements. Furthermore, the hardware makes mistakes in reading the state of the qubits, and finally there is a general shortage of computational quantum resources.\\

These opportunities and the challenges of the available quantum hardware forced us to set two main goals at the base of this work. The first goal is to use a gate-based quantum hardware to study the real-time dynamics of an SU(2) pure gauge theory on a small lattice in its Hamiltonian formulation. Secondly, to explore the available error mitigation techniques to mitigate the hardware error and extract a signal from noisy data.
\section{SU(2) pure gauge lattice theory \label{sec:S2}}
Our interest in the study of the SU(2) model resides in the fact that it is the smallest and simplest non-Abelian gauge theory. Of equal importance, it is also conceptually close to the Quantum Chromodynamics theory that is based on SU(3).
The SU(2) Hamiltonian in the common notation by Byrnes and Yamamoto \citep{Byrnes_Yamamoto} reads:
\begin{equation}\label{eq:H}
\hat H = \frac{g^2}{2}\left(\sum_{i={\rm links}}\hat E_i^2 - 2x\sum_{i={\rm plaquettes}}\hat\square_i\right) \,,
\end{equation}
where the only free parameter is the gauge coupling $g$, or the more convenient parameter $x\equiv2/g^4$. The first term in $\hat H$ is the chromoelectric field operator representing the stored chromoelectric energy on the lattice equal to $j(j+1)$ per each link. The second term is the chromomagnetic term that contains the plaquette operator $\hat\square_i$ made by the trace of the product of the 4 gauge links of the $i$th plaquette, and can add or subtract fluxes of energy on the lattice plaquettes.\\

\begin{figure}[H]
\centering
\includegraphics[width=0.7\linewidth]{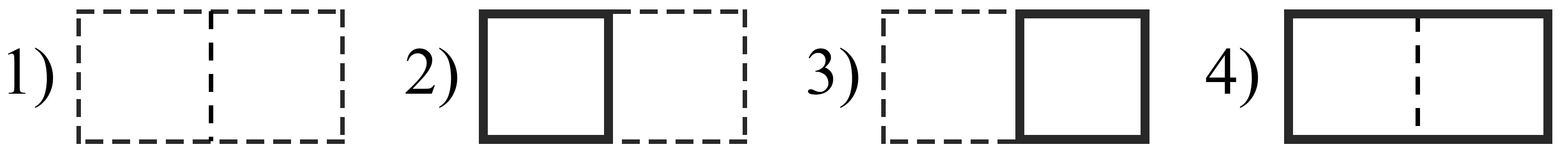}
\caption{The four possible states available on the 2-plaquette lattice after the $j_{max}=1/2$ truncation. Each solid line means a $j=1/2$ flux of energy is present, while the dashed line a $j=0$ meaning no energy is present. }
\label{fig:four_states}
\end{figure}

\begin{figure}[H]
\centering
\begin{subfigure}{.5\textwidth}
  \centering
  \includegraphics[width=1.\linewidth]{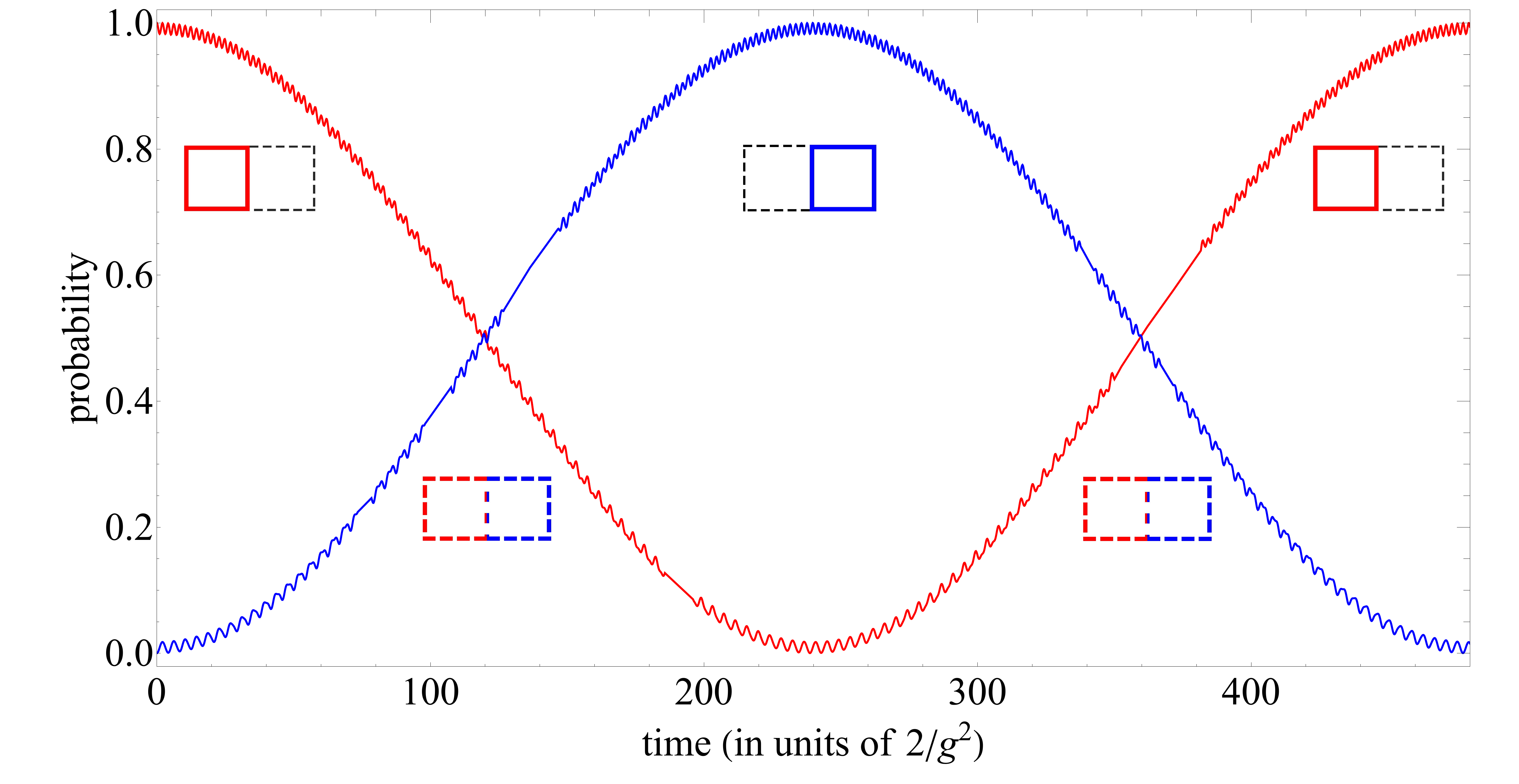}
  \label{fig:sub1}
\end{subfigure}%
\centering
\begin{subfigure}{.5\textwidth}
  \centering
  \includegraphics[width=1.\linewidth]{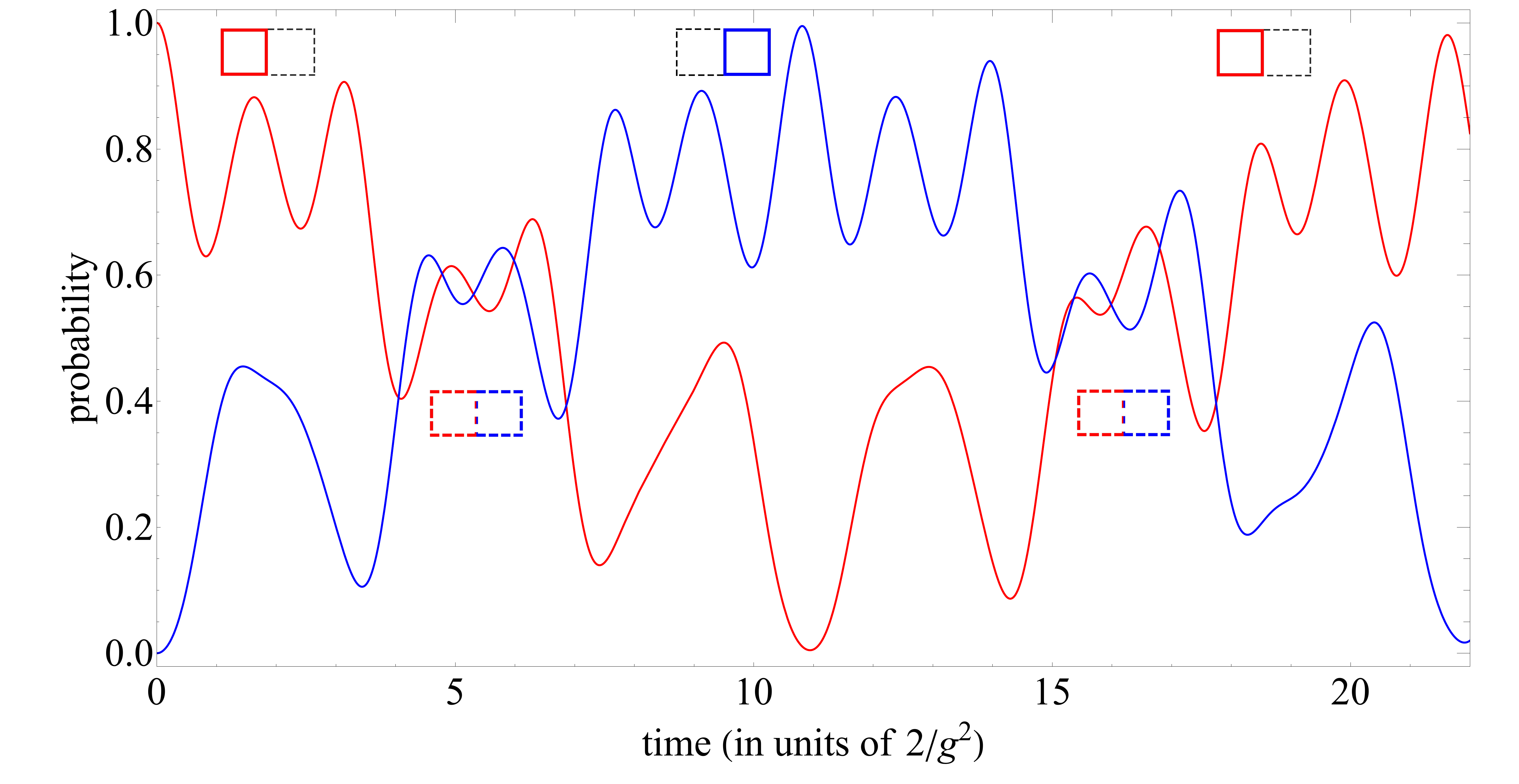}
  \label{fig:sub2}
\end{subfigure}
\caption{Exact calculation of the time evolution for the 2-plaquette lattice for $x<1$. The left panel is at $x=0.1$ while the right panel at $x=0.6$. The red (blue) solid curve is the exact probability of the left (right) plaquette having an energy flux of $j=1/2$ over time. In both panels the small diagrams using the notation of Fig.~\ref{fig:four_states} indicate the system state.}
\label{fig:Te_x<1}
\end{figure}

Only two lattices are considered in this work, both a single row with closed boundary conditions, but one with 2 plaquettes and the other with 5.
The energy spectrum of the theory is infinite, but considering the limitations of the current quantum hardware a truncation is necessary. Therefore we truncated the Hamiltonian allowing only the two lowest eigenstates for $j=0$ and $j=1/2$, so only four states are present, as shown in Fig.~\ref{fig:four_states}.
These four states are obtained by applying the plaquette operators to the vacuum state and obeying the SU(2) Gauss law. The interested reader can find the detailed formal derivation in previous work \citep{ARahman:2021ktn,PhysRevD.101.074512}. The truncated theory can be encoded on the quantum hardware using one qubit for each plaquette. Therefore the 4 states in Fig.~\ref{fig:four_states} can be represented with the following choice of qubits:
\begin{equation}\label{eq:states}
1) \longrightarrow \lvert 0 \rangle \lvert 0 \rangle,\qquad 2) \longrightarrow \lvert 0 \rangle \lvert 1 \rangle, \qquad  3) \longrightarrow \lvert 1 \rangle \lvert 0 \rangle, \qquad 4) \longrightarrow \lvert 1 \rangle \lvert 1 \rangle
\end{equation}
The truncated 2-plaquette Hamiltonian representation can be rewritten in gates as follows:
\begin{equation*} 
\begin{split}
\frac{2}{g^2}H = 
\left(
\begin{array}{cccc}
0 & -2x & -2x & 0 \\
-2x & 3 & 0 & -x \\
-2x & 0 & 3 & -x \\
0 & -x & -x & \frac{9}{2}
\end{array}
\right) &= \frac{3}{8}\left(7 - 3Z_0 - Z_0Z_1 - 3Z_1\right) - \frac{x}{2}(3+Z_1)X_0 - \frac{x}{2}(3+Z_0)X_1
\end{split}
\end{equation*}
The Hamiltonian has two regimes, one for large $x$ and one for small $x$.
For $x<1$ the Hamiltonian off-diagonal elements whose presence is due to the chromomagnetic operator are small. Therefore the dominant states are weakly coupled chromoelectric eigenstates. In this regime, the single-plaquette states will move across the lattice as shown in Fig.~\ref{fig:Te_x<1}.
A traveling exitation is the interpretation of the phenomena present in the left panel of Fig.~\ref{fig:Te_x<1} where for $x<1$ the probability of having an energy flux in the left plaquette (red curve) drops from $100\%$ at time $t=0$ to close to 0$\%$ at time $t=240$ while the probability of right plaquette (blue curve) does the opposite. Similarly in the right panel, the now larger value of $x$ superimposes some high frequencies.
That the excitation propagation time is shorter for larger $x$ is clearly visible by comparing the pseudo-period of the left panel at $x=0.1$ with the one present in the right panel at $x=0.6$.

\begin{figure}[H]
\centering
\begin{subfigure}{.5\textwidth}
  \centering
  \includegraphics[width=1.0\linewidth]{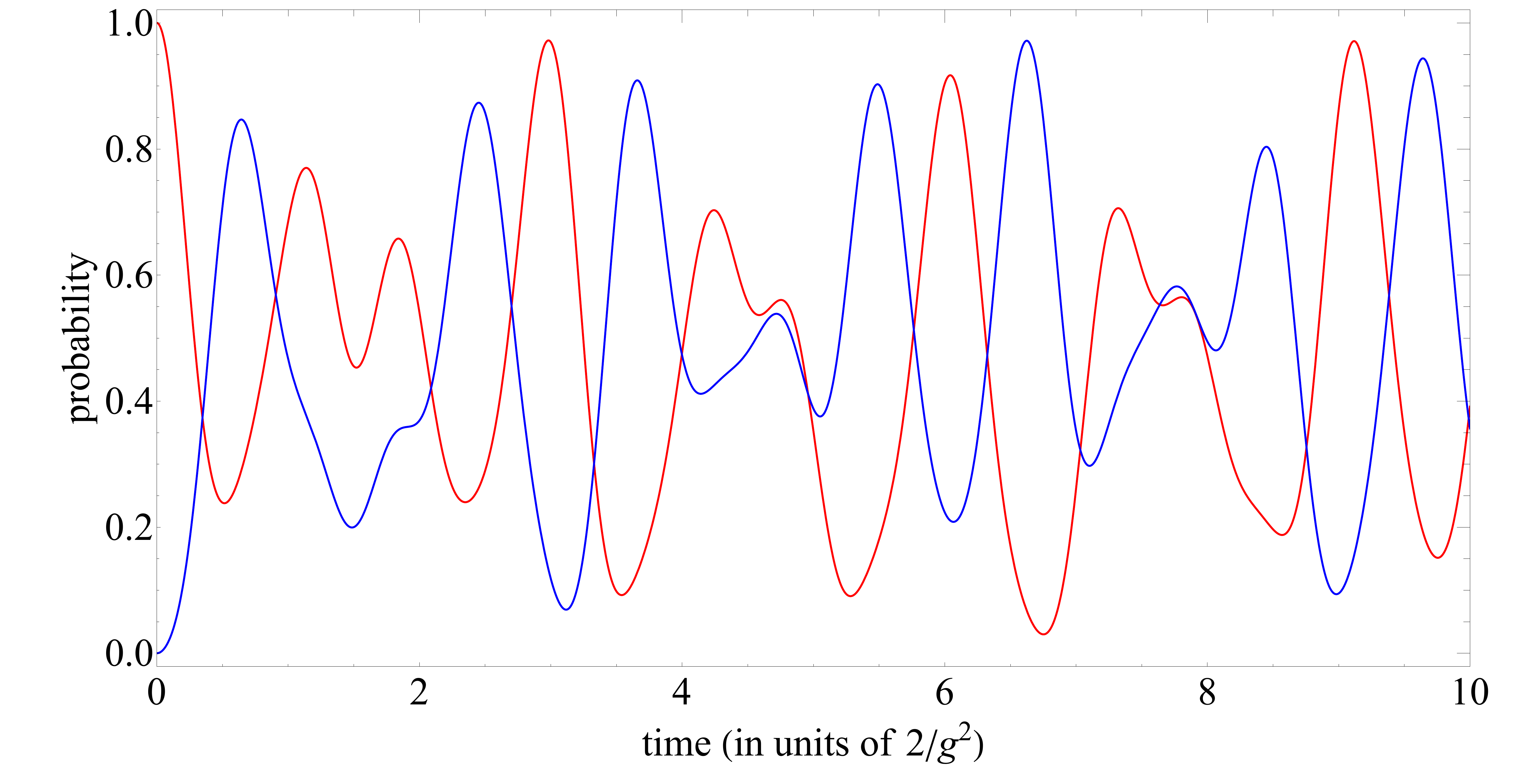}
  \label{fig:sub1}
\end{subfigure}%
\begin{subfigure}{.5\textwidth}
  \centering
  \includegraphics[width=1.0\linewidth]{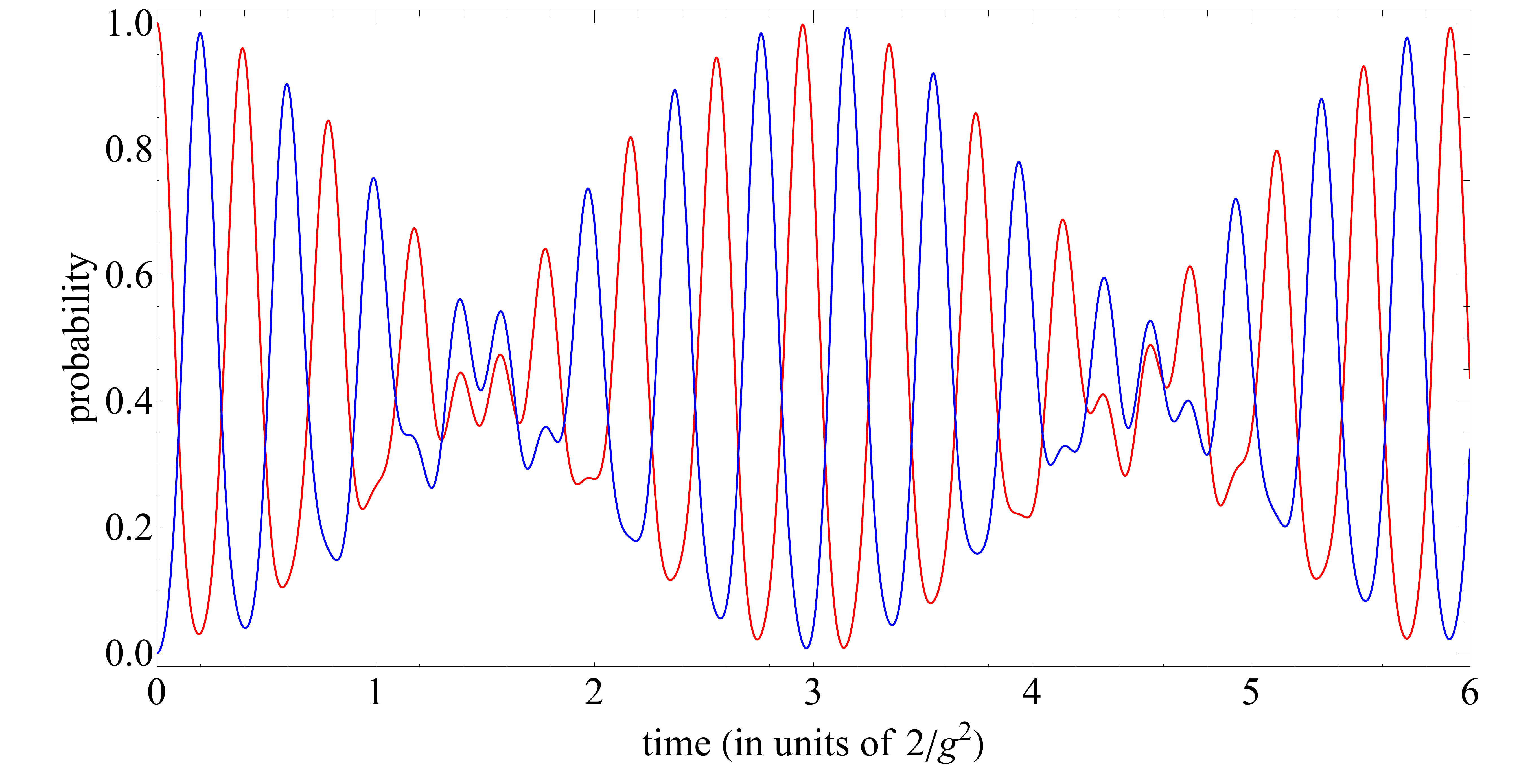}
  \label{fig:sub2}
\end{subfigure}
\caption{Exact calculation of the time evolution for the 2-plaquette lattice for $x>1$. The left panel is at $x=1.5$ while the right panel is at $x=5.0$. The red (blue) solid curve is the exact probability of the left (right) plaquette having an energy flux of $j=1/2$ over time.}
\label{fig:Te_x>1}
\end{figure}

For $x>1$ the chromomagnetic contribution dominates mixing of the single-plaquette states. This is evident by the presence of many higher frequencies superimposed to the single-plaquettes transitions in Fig.~\ref{fig:Te_x>1}.

At larger values of $x$, higher frequencies are present as can be seen moving from the left panel at $x = 1.5$ to the right panel at $x = 5.0$ in Fig.~\ref{fig:Te_x>1}.\\

A python code running inside the IBMQ quantum lab \citep{IBM_LAB} accesses the quantum resources and implements the time evolution operator and the needed error mitigation techniques. These are both described in the following sections.

\section{Time evolution on the quantum hardware}
The time evolution of the SU(2) theory can be studied using a quantum computer once the time evolution operator $e^{-iHt}$ has been written in gates. This can be done by approximating the time evolution operator:
\begin{equation}\label{eq:states}
e^{-i\left( \frac{3}{8}\left(7 - 3Z_0 - Z_0Z_1 - 3Z_1\right) - \frac{x}{2}(3+Z_1)X_0 - \frac{x}{2}(3+Z_0)X_1\right) t}
\end{equation}
for small time step $dt$ through the use of the second order Suzuki-Trotter expansion \citep{Hatano2005}:
\begin{equation*}\label{eq:ST_expansion}
e^{-iHt} = e^{-i\sum_{j=1}^m H_jt}=\left(  \prod_{j=1}^m  e^{-i H_j \, dt/2} \: \prod_{j=m}^1  e^{-i H_j\, dt/2} \right)^{N_t}    + O\left( m^2 t \, dt \right) 
\end{equation*}

\begin{figure}[H]
\centering
\includegraphics[width=0.5\linewidth]{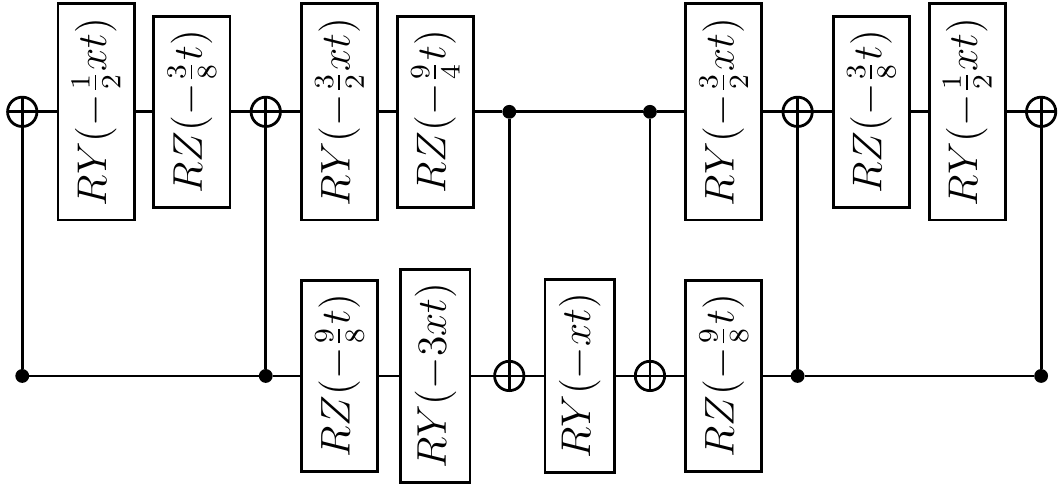}
\caption{A single second-order Suzuki-Trotter step for the 2-plaquette lattice. }
\label{fig:circuit_TE2pla}
\end{figure}

A single second-order Suzuki-Trotter step written in gates requires 6 CNOT gates as shown in Fig.~\ref{fig:circuit_TE2pla} that are further reduced to 4 in view of cancellations after combination with neighbouring time steps.

\begin{figure}[H]
\centering
\includegraphics[width=0.8\linewidth]{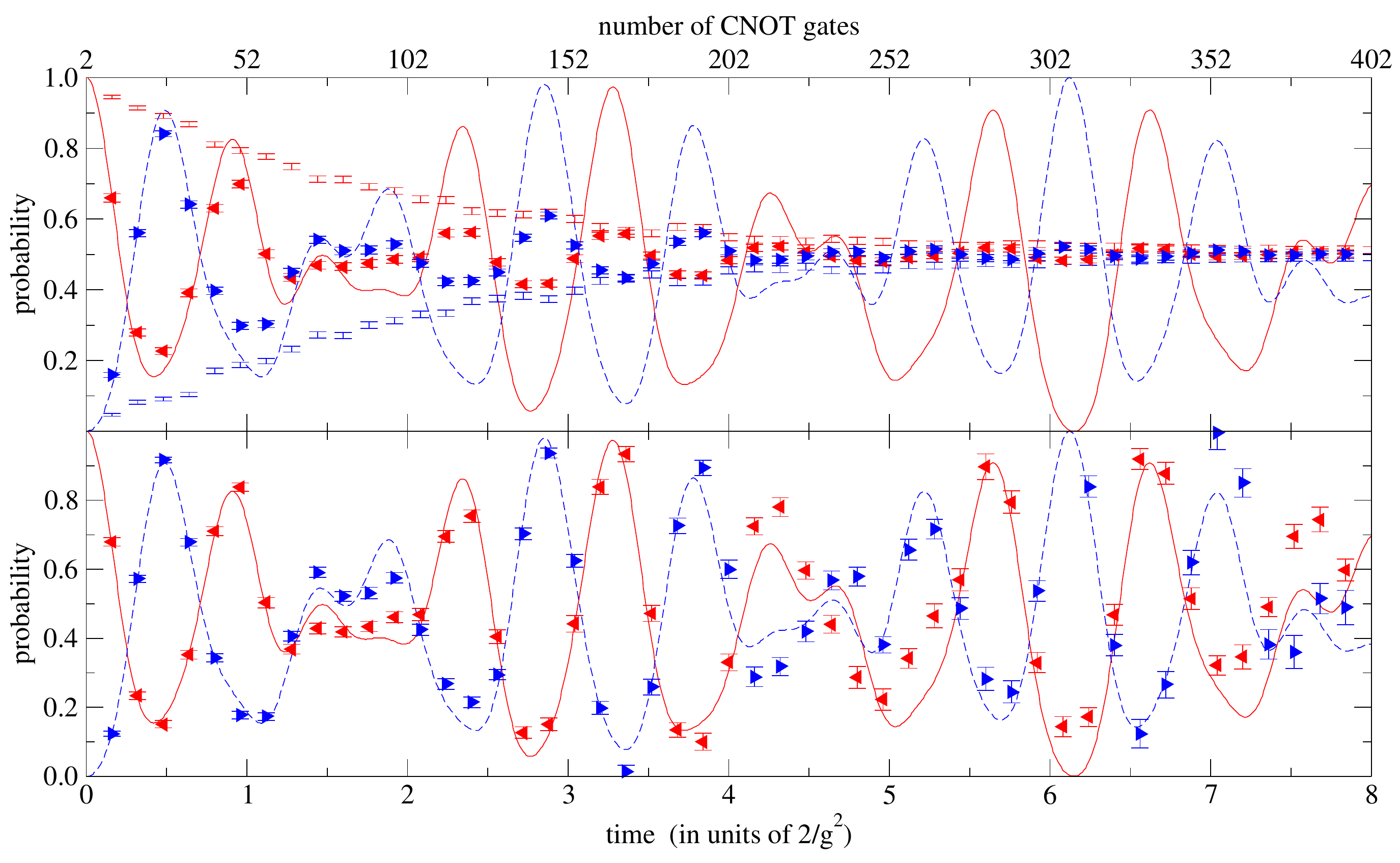}
\caption{Real-time evolution of the state with only a $j=1/2$ flux of energy in the left plaquette for the 2-plaquette lattice case at gauge coupling $x=2.0$ and time step $dt=0.08$.  
The red solid (blue dashed) curve is the exact probability of the left (right) plaquette having an energy flux of $j=1/2$ over time.  In the upper panel: The red left-pointing (blue right-pointing) triangles are the physics data from runs on the quantum hardware {\tt ibm\_lagos}. The red (blue) error bars without symbols are the error mitigation data from runs on {\tt ibm\_lagos}. In the lower panel the triangles are the final results of our error mitigation method that processes the upper panel data through the self-mitigation equation Eq.~(\ref{ed:SFM_EQ}). 
}
\label{fig:TE2plagate_x2}
\end{figure}

The cancellations of the edge CNOT gates from neighbouring time steps and the combination of the edge rotation gates is done at code level to minimize the total number of gates.
The time evolution circuit can be submitted to the quantum hardware to study how a chosen initial state evolves in real time as shown in Fig.~\ref{fig:TE2plagate_x2}.

The graph in  Fig.~\ref{fig:TE2plagate_x2} presents the result from studying the real-time evolution of the 2-plaquette system starting in the state with only a $j=1/2$ flux of energy in the left plaquette. In the upper panel the data represented by triangles are obtained by running the time evolution circuits together with two error mitigation techniques, the mitigation of measurement error and randomized compiling. The error bars without symbol are the results of the error-estimation circuits needed for our self-mitigation method.  In the lower panel the final result is obtained by processing the upper panel data with self-mitigation.\\

The two main features are: the long-time range that spans dozens of time steps with up to 400 CNOT gates, and the ability of our error mitigation method to extract a signal from noisy data as shown in the lower panel. For $t>4$ the signal in the upper panel appears to be dominated by noise but in the corresponding lower panel, where the points have been error mitigated with self-mitigation, they are corrected and close to the exact result. The final agreement with the exact curve is not perfect, but only a small discrepancy remains.\\

In the following sections all the error mitigation techniques and the measurement protocol used in this work are described.

\section{Error mitigation techniques}

The common available quantum computers are noisy. The imperfect realization gives rise to results littered with errors limiting their usefulness. To extend the usability and extract value from the existing NISQ hardware, researchers are developing and using methods to reduce the hardware error. 
Even if the time evolution study is a conceptual, simple process its circuit realization requires the use of many gates, in particular many CNOT gates that are the noisiest, therefore a comprehensive set of error mitigation techniques is needed:

\begin{enumerate}

\item	Mitigation of measurement error \citep{PhysRevA.103.042605}:\\
This is a procedure used to correct the errors that the hardware makes in measuring the qubit's state.
If we are working with $n$ qubits, there are $2^n$ possible pure states in which the result of the measurement of the system can be found. 
Each of these $2^n$ states have to be recreated on the hardware with a dedicated circuit and measured to construct the $2^n\times2^n$ calibration matrix, whose entries are the probabilities that a particular state, once measured, has a superposition with another state.
In case of no hardware errors, for example on the simulator, the calibration matrix is diagonal, meaning that the states have no superposition with each other.
In the case of a 2-plaquette lattice we use 2 qubits, therefore there are 4 mitigation circuits as shown in Fig.~\ref{fig:4_circuits}. 

\begin{figure}[H]
\centering
\includegraphics[width=0.9\linewidth]{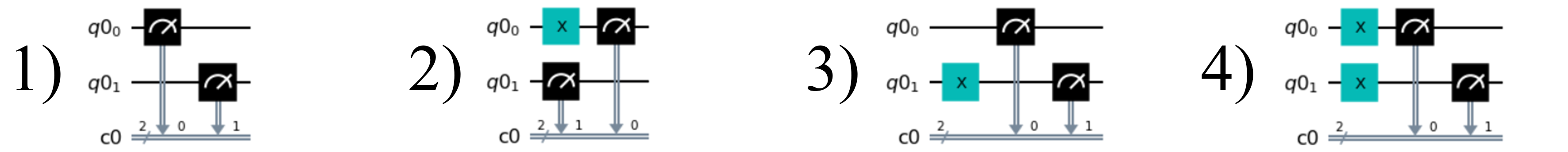}
\caption{The four circuits needed by the mitigation of measurement error for the 2-plaquette case.  }
\label{fig:4_circuits}
\end{figure}

Finally the calibration matrix is then applied through a fitting procedure called sequential least squares programming to mitigate the measurement error present in the physical circuit.

\item	Randomized compiling \citep{Wallman_2016} (Pauli-twirling):\\
This is a technique used to transform the CNOT coherent noise into incoherent noise. This is done by creating new physics circuits in which each original CNOT gate is equivalently substituted by a CNOT gate “surrounded” by Pauli or identity gates. It can be shown that there are 16 possible ways to rewrite a CNOT gate as shown in Fig.~\ref{fig:randomized_c}:
\begin{figure}[H]
\centering
\includegraphics[width=0.9\linewidth]{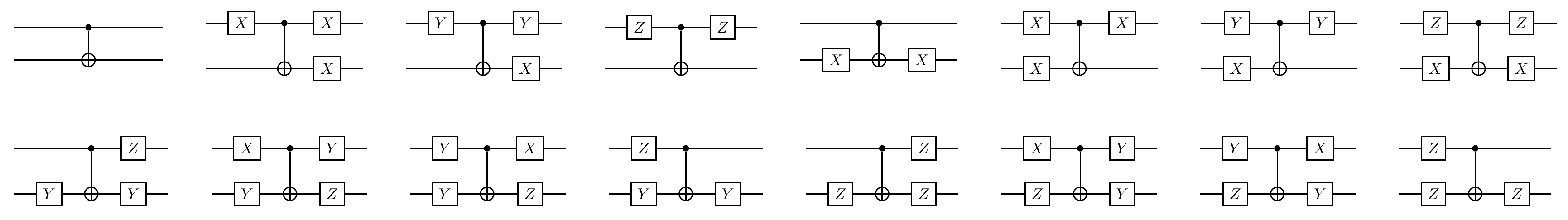}
\caption{The 16 possible ways to rewrite a CNOT gate using combinations of Pauli gates and the identity.}
\label{fig:randomized_c}
\end{figure}
Therefore, the physics circuit has to be run a sufficient number of times to randomly access many possible combinations for the surrounding gate of each CNOT gate in the original physics circuit. The final physics measurement is obtained by averaging the measurement of these runs.

\item	Zero-noise extrapolation \citep{Li_Benjamin}:\\
The method consists to study the circuit noise by creating copies of the original circuits where the noise is artificially increased by replacing each CNOT gate by odd multiples (triplet, quintet etc.) as shown in Fig.~\ref{fig:Cnot_triplet}.

\begin{figure}[H]
\centering
\begin{subfigure}{.5\textwidth}
  \centering
  \includegraphics[width=0.6\linewidth]{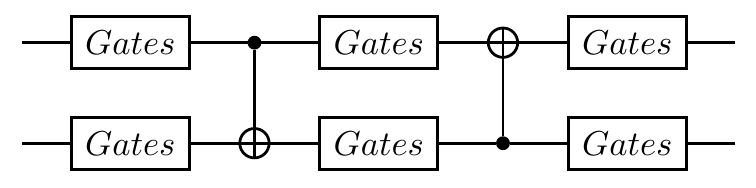}
  \label{fig:sub1}
\end{subfigure}%
\begin{subfigure}{.50\textwidth}
  \centering
  \includegraphics[width=0.9\linewidth]{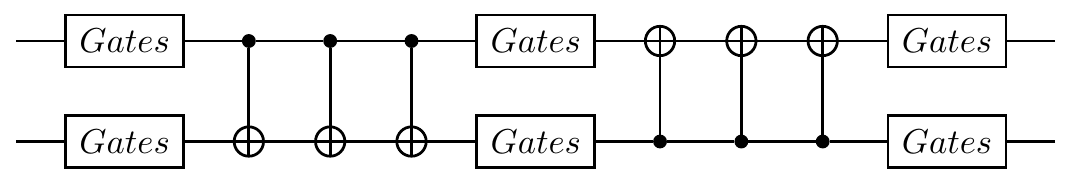}
  \label{fig:sub2}
\end{subfigure}
\caption{On the left the original circuit, on the right the same circuit in which the CNOT gates are replaced by a CNOT triplet.}
\label{fig:Cnot_triplet}
\end{figure}

The zero-noise limit is extracted by fitting the circuit's measurements with a function of the CNOT multiples, as the intercept of the curve with the y-axis.

\item	Self-mitigation:\\
This method belongs to a class of error mitigation techniques in which an extra circuit is used to estimate the hardware errors. The error estimation circuit should be as similar as possible to the gate structure of the physics circuit and should have a known exact result. The estimation of the hardware error from the error estimation circuits is used to mitigate the hardware error on the physics circuit.

With the self-mitigation method, the error estimation circuit is identical to the physics circuit.
\begin{figure}[H]
\centering
\includegraphics[width=0.6\linewidth]{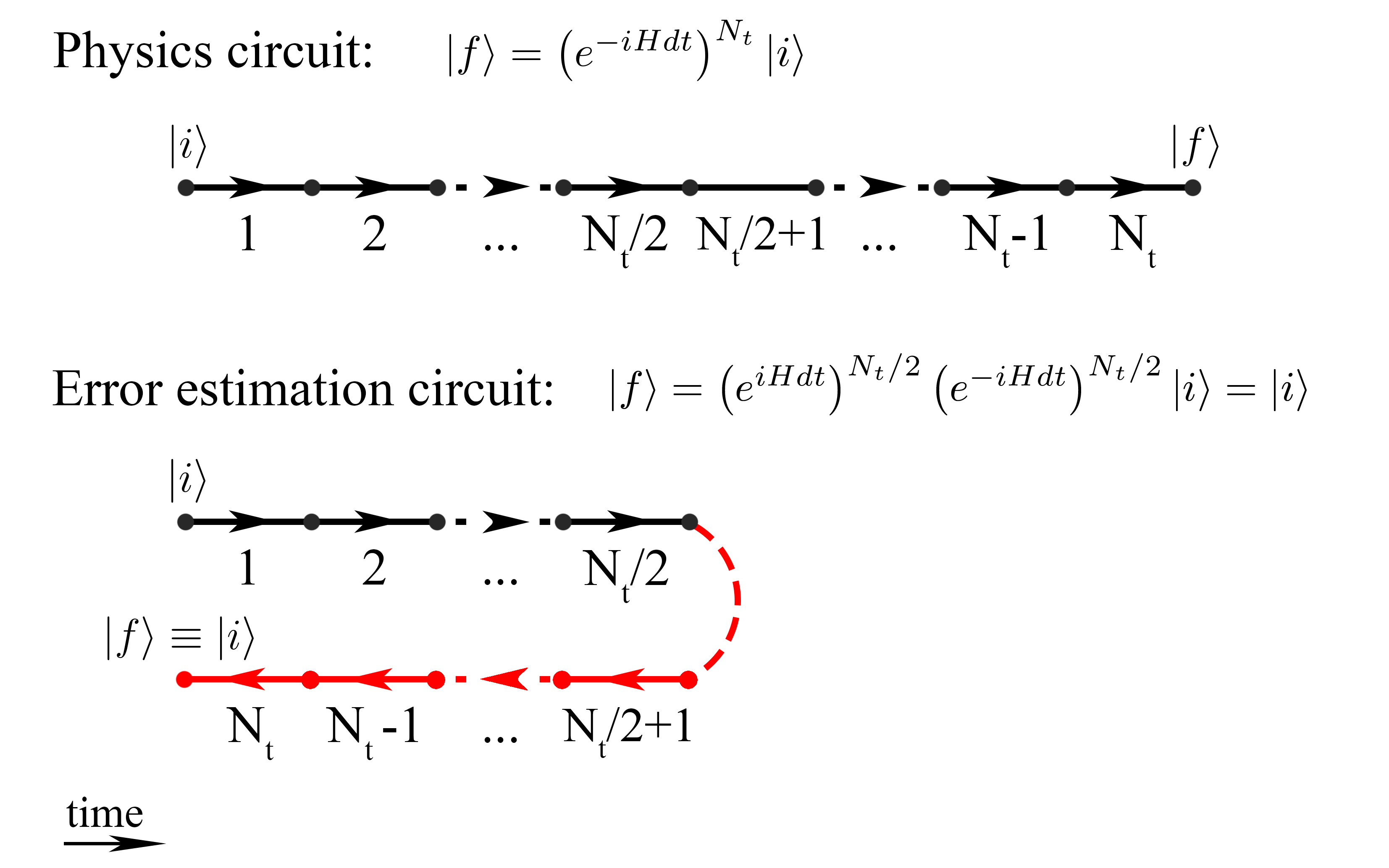}
\caption{Schematic representation of how the time evolution steps are organized in the error mitigation circuits versus the physics circuit.}
\label{fig:Fig_self_mit}
\end{figure}

As shown in Fig.~\ref{fig:Fig_self_mit} the physics circuit does $N$ time steps forward in time. In comparison the error estimation circuit does half time steps forward and half backward, returning to the initial state.
Therefore, the error estimation circuit has the same gates in the same order and the identical or opposite variable inside the rotation gates, and accordingly it closely reproduces the physics circuit noise.
The hardware error can be estimated in the disparity between the final state of the error estimation circuit and the initial state.
Due to the variability of the hardware the two circuits have to run back-to-back to ensure the hardware noise affecting them is similar (if not identical). Therefore the error mitigated final result for the physics circuit is obtained through the self-mitigation equation:
\begin{equation}
\left.\frac{P_{\rm true}-\tfrac{1}{2}}{P_{\rm computed}-\tfrac{1}{2}}\right|_{\rm physics\,run}
=
\left.\frac{P_{\rm true}-\tfrac{1}{2}}{P_{\rm computed}-\tfrac{1}{2}}\right|_{\rm mitigation\,run}
\label{ed:SFM_EQ}
\end{equation}

\end{enumerate}

\section{Measurement protocol}
Using the 7 qubits quantum hardware {\tt ibm\_lagos} allows one to run a sequence of 300 circuits back to back we chose $10^4$ hits. To further clarify our approach, the protocol used is as follows:  
\begin{enumerate}

\item	Submit the $2^n$ circuits for the mitigation of measurement error.
(4 circuits for the 2-plaquette and 32 circuits for the 5-plaquette case.)
\item	Submit the randomized compiling circuits for the error estimation circuit and the physics circuit.
(148 circuits for the error estimation circuits and 148 for the physics circuits for the 2-plaquette case. 134 error estimation circuits and 134 physics circuits for the 5-plaquette case.)
\item	Collect all the measurements.
\item	Apply the measurement-error calibration matrix to the error estimation circuit and the physics circuit results.
\item	Use the self-mitigation equation to mitigate the hardware error and extract the mitigated value of the measurements. 
\item	Calculate the final error for the error estimation and the physics results as the sum in quadrature of the statistical error from the $10^4$ hits and the error from the bootstrap samples. 

\end{enumerate}

\section{Numerical Results}
The developed method can be used to study interesting features of the theory, for example the traveling excitation across the 2-plaquette lattice. As we already described in Sec.~\ref{sec:S2} to observe traveling excitation we need to consider a value of the gauge coupling $x<1$, where the chromoelectric part dominates and therefore the single-plaquette states are the main contributor.

\begin{figure}[H]
\centering
\includegraphics[width=0.6\linewidth]{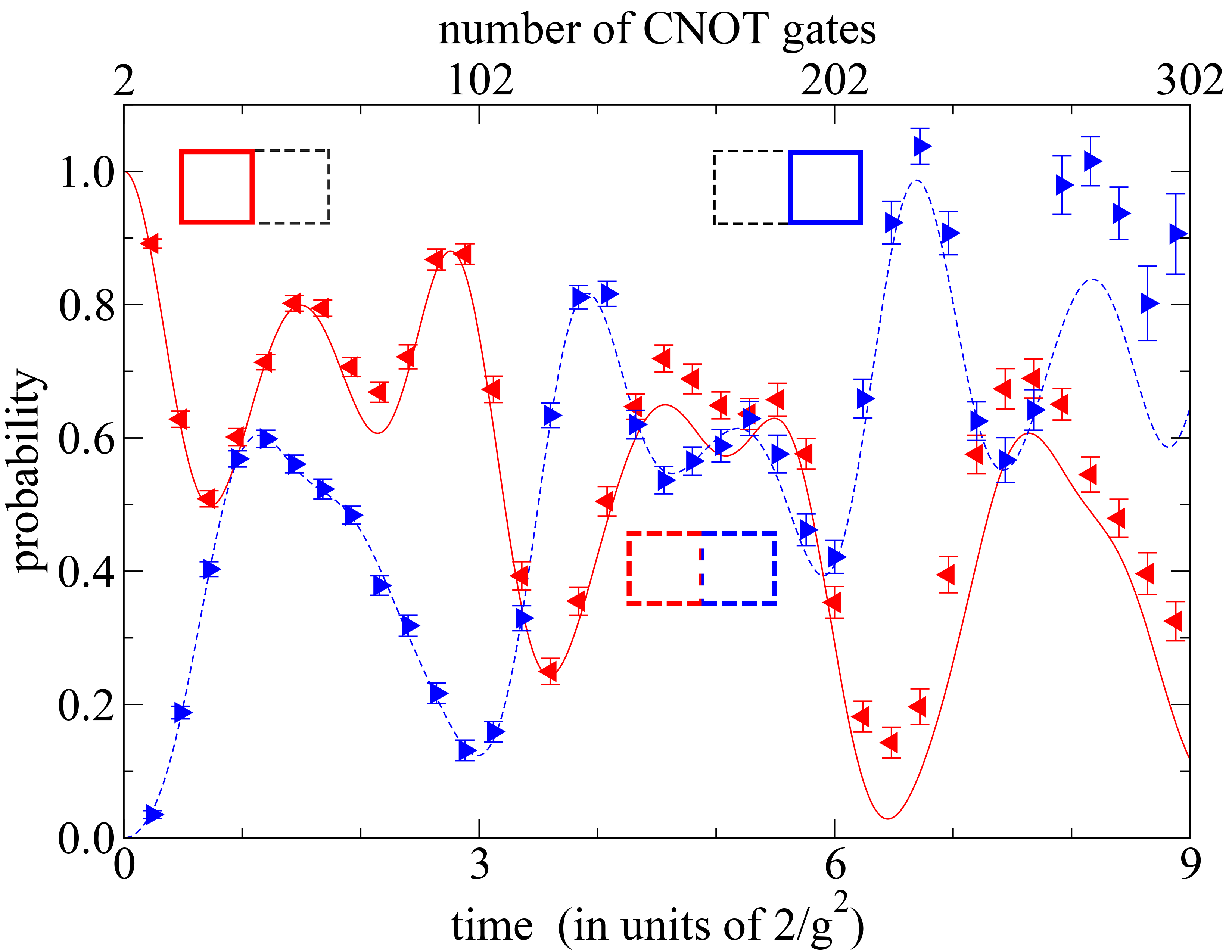}
\caption{Real-time excitation obtained as an evolution of the state with only a $j=1/2$ flux of energy in the left plaquette for the 2-plaquette lattice case at gauge coupling $x=0.8$ and 2 time steps per point with $dt=0.12$.  The red solid (blue dashed) curve is the exact probability of the left (right) plaquette having an energy flux of j=1/2 over time. The red left-pointing (blue right-pointing) triangles are the physics data from calculations done on the quantum hardware {\tt ibm\_lagos} and processed by self-mitigation.}
\label{fig:TE_2plaTE}
\end{figure}

The physical interpretation of the phenomenon can be clearer if we interpret the closed loop with $j=1/2$ that is a gauge invariant excitation as an approximation of a SU(2) glueball. Therefore, the traveling excitation is like a particle moving across the lattice from left to right.\\

The analytical method and measurement protocol can be extended to a larger lattice. On the 7 qubits hardware {\tt ibm\_lagos}, the longest qubit chain with nearest-neighbour connectivity has 5 qubits, therefore avoiding the use of swap gates and using 1 qubit for each plaquette, the largest implementable lattice has 5 plaquettes.

\begin{figure}[H]
\centering
\includegraphics[width=0.6\linewidth]{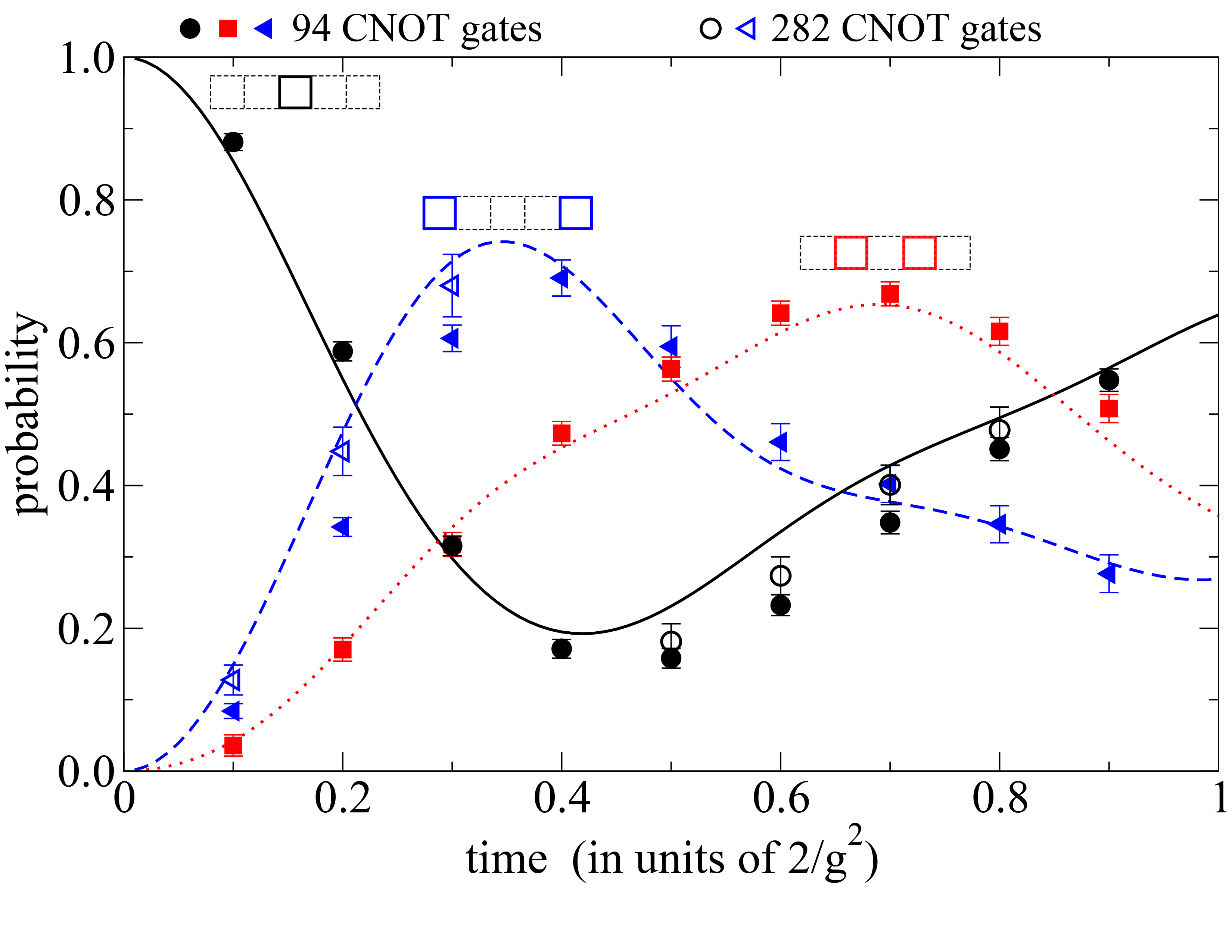}
\caption{Real-time evolution of the state with only a $j=1/2$ flux of energy in the left plaquette for the 5-plaquette lattice case at gauge coupling $x=2.0$ and 4 Trotter steps per point at different values of $dt$. The black solid, blue dashed, and the red dotted curves are respectively the exact probability that the flux of energy $j=1/2$ is in the centre plaquette, the two edge plaquettes and the two neighbours to the centre plaquette. The solid black circle, blue triangle, and red square are the corresponding data from the quantum hardware {\tt ibm\_lagos} after self-mitigation.  The open circle, triangle, and square are quantum data from {\tt ibm\_lagos} after using both self-mitigation and zero-noise extrapolation.
}
\label{fig:TE_5plagate}
\end{figure}

The results in Fig.~\ref{fig:TE_5plagate} are obtained from 4 jobs each with 134 runs using 4 second-order Trotter steps with different step size and totalling 94 CNOT gates for each point. The agreement is reasonable but not perfect as can be seen from the comparison of the filled data points with the exact curves, and can be improved as shown by the empty symbols using the zero-noise extrapolation using up to a CNOT triplet. In contrast to Fig.~\ref{fig:TE_2plaTE}, the curves cannot be interpreted as a traveling excitation because at the value used for the gauge coupling x=2.0, the chromomagnetic contribution dominates.

\section{Summary}
This work demonstrates that an IBM gate-based quantum computer can be successfully used to study the real-time evolution of a non-Abelian lattice gauge theory on a small lattice. As shown in Fig.~\ref{fig:TE2plagate_x2}  the real-time evolution study of a SU(2) theory was extended to a time range much larger than previous studies on non-Abelian gauge theories \citep{PhysRevD.101.074512,ARahman:2021ktn,PhysRevD.103.094501,Illa:2022jqb}. Furthermore, in Fig.~\ref{fig:TE_2plaTE}  we reported the observation of an example of a real-time dynamical process: a SU(2) local excitation moving across the lattice.
To obtain these results we used well-known error mitigation techniques such as measurement mitigation and randomized compiling. By themselves, these error mitigation techniques are not sufficient. To extract signal at large-time, we introduced a new approach called self-mitigation, which uses the same physics circuit as its own noise-mitigation circuit.
In Fig.~\ref{fig:TE_5plagate}  we not only show that the current quantum hardware can be used for the time evolution study of a larger lattice, but that self-mitigation can be successfully combined with the mitigation method called zero-noise extrapolation to increase the error mitigation. Self-mitigation was introduced in \citep{Rahman:2022rlg} and was used in \citep{Atas:2022dqm}. A similar approach was used in \citep{Farrell:2022wyt}.\\

An interested reader may find the article by the same authors useful \citep{Rahman:2022rlg}, as the original Lattice 2022 talk was based on it.

\section{Acknowledgements}
The present work was supported in part by a Discovery Grant from the Natural Sciences and Engineering Research Council (NSERC) of Canada and by an NSERC Undergraduate Student Research Award. The authors acknowledge the use of IBM Quantum services for this work. The views expressed are those of the authors, and do not reflect the official policy or position of IBM or the IBM Quantum team.

\bibliographystyle{JHEP}
\setlength{\bibsep}{3pt plus 0.3ex}
\bibliography{PoS2022_Bib.bib}

\providecommand{\href}[2]{#2}\begingroup\raggedright\begin{thebibliography}{10}

\bibitem{JohnPreskillNISQ}
J.~Preskill, \emph{Simulating quantum field theory with a quantum computer},
  \href{https://arxiv.org/abs/1811.10085}{{\ttfamily 1811.10085}}.

\bibitem{Aharonov_et_all}
D.~Aharonov and M.~Ben-Or, \emph{Fault-tolerant quantum computation with
  constant error rate},
  \href{https://doi.org/10.1137/S0097539799359385}{\emph{SIAM Journal on
  Computing} {\bfseries 38} (2008) 1207}
  [\href{https://arxiv.org/abs/quant-ph/9906129}{{\ttfamily
  quant-ph/9906129}}].

\bibitem{Bauer:2022hpo}
C.W.~Bauer et~al., \emph{{Quantum Simulation for High Energy Physics}},
  \href{https://arxiv.org/abs/2204.03381}{{\ttfamily 2204.03381}}.

\bibitem{Zi-Xiang_Hong}
Z.-X.~Li and H.~Yao, \emph{{Sign-Problem-Free fermionic quantum Monte Carlo:
  Developments and applications}},
  \href{https://doi.org/10.1146/annurev-conmatphys-033117-054307}{\emph{Annual
  Review of Condensed Matter Physics} {\bfseries 10} (2019) 337}
  [\href{https://arxiv.org/abs/1805.08219}{{\ttfamily 1805.08219}}].

\bibitem{Sign_problem_Gattringer}
C.~Gattringer and K.~Langfeld, \emph{{Approaches to the sign problem in lattice
  field theory}}, \href{https://doi.org/10.1142/S0217751X16430077}{\emph{Int.
  J. Mod. Phys. A} {\bfseries 31} (2016) 1643007}
  [\href{https://arxiv.org/abs/1603.09517}{{\ttfamily 1603.09517}}].

\bibitem{Byrnes_Yamamoto}
T.~Byrnes and Y.~Yamamoto, \emph{Simulating lattice gauge theories on a quantum
  computer}, \href{https://doi.org/10.1103/PhysRevA.73.022328}{\emph{Phys. Rev.
  A} {\bfseries 73} (2006) 022328}
  [\href{https://arxiv.org/abs/quant-ph/0510027}{{\ttfamily
  quant-ph/0510027}}].

\bibitem{ARahman:2021ktn}
S.~A~Rahman, R.~Lewis, E.~Mendicelli and S.~Powell, \emph{{SU(2) lattice gauge
  theory on a quantum annealer}},
  \href{https://doi.org/10.1103/PhysRevD.104.034501}{\emph{Phys. Rev. D}
  {\bfseries 104} (2021) 034501}
  [\href{https://arxiv.org/abs/2103.08661}{{\ttfamily 2103.08661}}].

\bibitem{PhysRevD.101.074512}
N.~Klco, M.J.~Savage and J.R.~Stryker, \emph{{SU(2) non-Abelian gauge field
  theory in one dimension on digital quantum computers}},
  \href{https://doi.org/10.1103/PhysRevD.101.074512}{\emph{Phys. Rev. D}
  {\bfseries 101} (2020) 074512}
  [\href{https://arxiv.org/abs/1908.06935}{{\ttfamily 1908.06935}}].

\bibitem{IBM_LAB}
``Ibm quantum lab.'' http://quantum-computing.ibm.com/.

\bibitem{Hatano2005}
N.~Hatano and M.~Suzuki, \emph{Finding exponential product formulas of higher
  orders},  in \emph{Quantum Annealing and Other Optimization Methods}, A.~Das
  and B.~K.~Chakrabarti, eds., (Berlin, Heidelberg), pp.~37--68, Springer
  Berlin Heidelberg (2005), \href{https://doi.org/10.1007/11526216_2}{DOI}
  [\href{https://arxiv.org/abs/math-ph/0506007}{{\ttfamily math-ph/0506007}}].

\bibitem{PhysRevA.103.042605}
S.~Bravyi, S.~Sheldon, A.~Kandala, D.C.~Mckay and J.M.~Gambetta,
  \emph{Mitigating measurement errors in multiqubit experiments},
  \href{https://doi.org/10.1103/PhysRevA.103.042605}{\emph{Phys. Rev. A}
  {\bfseries 103} (2021) 042605}
  [\href{https://arxiv.org/abs/2006.14044}{{\ttfamily 2006.14044}}].

\bibitem{Wallman_2016}
J.J.~Wallman and J.~Emerson, \emph{Noise tailoring for scalable quantum
  computation via randomized compiling},
  \href{https://doi.org/10.1103/physreva.94.052325}{\emph{Physical Review A}
  {\bfseries 94} (2016) } [\href{https://arxiv.org/abs/1512.01098}{{\ttfamily
  1512.01098}}].

\bibitem{Li_Benjamin}
Y.~Li and S.C.~Benjamin, \emph{Efficient variational quantum simulator
  incorporating active error minimization},
  \href{https://doi.org/10.1103/PhysRevX.7.021050}{\emph{Phys. Rev. X}
  {\bfseries 7} (2017) 021050}
  [\href{https://arxiv.org/abs/1611.09301}{{\ttfamily 1611.09301}}].

\bibitem{PhysRevD.103.094501}
A.~Ciavarella, N.~Klco and M.J.~Savage, \emph{{Trailhead for quantum simulation
  of SU(3) Yang-Mills lattice gauge theory in the local multiplet basis}},
  \href{https://doi.org/10.1103/PhysRevD.103.094501}{\emph{Phys. Rev. D}
  {\bfseries 103} (2021) 094501}
  [\href{https://arxiv.org/abs/2101.10227}{{\ttfamily 2101.10227}}].

\bibitem{Illa:2022jqb}
M.~Illa and M.J.~Savage, \emph{Basic elements for simulations of standard model
  physics with quantum annealers: Multigrid and clock states},
  \href{https://arxiv.org/abs/2202.12340}{{\ttfamily 2202.12340}}.

\bibitem{Rahman:2022rlg}
S.A.~Rahman, R.~Lewis, E.~Mendicelli and S.~Powell, \emph{{Self-mitigating
  Trotter circuits for SU(2) lattice gauge theory on a quantum computer}},
  \href{https://doi.org/10.1103/PhysRevD.106.074502}{\emph{Phys. Rev. D}
  {\bfseries 106} (2022) 074502}
  [\href{https://arxiv.org/abs/2205.09247}{{\ttfamily 2205.09247}}].

\bibitem{Atas:2022dqm}
{Y.~Y.~Atas, J.~F.~Haase, J.~Zhang, V.~Wei, S.~M.~L. Pfaendler, R.~Lewis and
  C.~A.~Muschik}, \emph{{Real-time evolution of SU(3) hadrons on a quantum
  computer}},  \href{https://arxiv.org/abs/2207.03473}{{\ttfamily 2207.03473}}.

\bibitem{Farrell:2022wyt}
R.C.~Farrell, I.A.~Chernyshev, S.J.M.~Powell, N.A.~Zemlevskiy, M.~Illa and
  M.J.~Savage, \emph{{Preparations for Quantum Simulations of Quantum
  Chromodynamics in 1+1 Dimensions: (I) Axial Gauge}},
  \href{https://arxiv.org/abs/2207.01731}{{\ttfamily 2207.01731}}.

\end{thebibliography}\endgroup
\end{document}